\documentclass{aastex}
\usepackage{apjfonts}

\slugcomment{Apr.\ 24, 2001}

\shorttitle{Mass--Metallicity Relation for Dwarf Galaxies}
\shortauthors{Tamura, Hirashita, \& Takeuchi}

\begin{document}

\title{Mass--Metallicity Relation for the Local Group Dwarf
Spheroidal Galaxies: A New Picture for the Chemical Enrichment
of Galaxies in the Lowest Mass Range}

\author{Naoyuki Tamura and Hiroyuki Hirashita\altaffilmark{1}}
\affil{Department of Astronomy, Faculty of Science, Kyoto University,
Sakyo-ku, Kyoto 606--8502, Japan}
\email{tamura, hirasita@kusastro.kyoto-u.ac.jp}
\and
\author{Tsutomu T. Takeuchi}
\affil{Division of Particle and Astrophysical Sciences, School of Science, 
Nagoya University, Chikusa-ku, Nagoya 464--8602, Japan}
\email{takeuchi@u.phys.nagoya-u.ac.jp}

\altaffiltext{1}{Research Fellow of the Japan Society for the Promotion
of Science.}

\begin{abstract}

The virial mass ($M_{\rm vir}$)--metallicity relation among the Local
Group dwarf spheroidal galaxies (dSphs) is examined.
Hirashita, Takeuchi, \& Tamura showed that the dSphs can be divided into
two distinct classes with respect to the relation between their virial
masses and luminosities: low-mass ($M_{\rm vir} \la 10^8 M_\odot$) and
high-mass ($M_{\rm vir} \ga 10^8 M_\odot$) groups.
We see that both the mass--metallicity and the mass--luminosity
relations of the high-mass dSphs are understood as a low-mass extension
of giant ellipticals.
On the contrary, we find that the classical galactic-wind model is
problematic to apply to the low-mass dSphs, whose low binding energy is
comparable to that released by several supernova explosions.
A strongly regulated star formation in their formation phase is required
to reproduce their observed metallicity. Such regulation is naturally
expected in a gas cloud with the primordial elemental abundance
according to Nishi \& Tashiro.
A significant scatter in the mass--metallicity relation for the low-mass
dSphs is also successfully explained along with the scenario of
Hirashita and coworkers. We not only propose a new picture for a
chemical enrichment of the dSphs, but also suggest that the
mass--metallicity and the mass--luminosity relations be understood in a
consistent context.

\end{abstract}


\section{INTRODUCTION}

Dwarf spheroidal galaxies (dSphs) are faint galaxies, whose luminosity
$L$ is $10^5$--$10^7\, L_\odot$.
They are characterized by their low surface brightnesses (Gallagher
\& Wyse 1994 for a review).
The studies of such low-luminosity objects are important for several
reasons. For example, since lower-mass galaxies are considered to have
formed earlier in a hierarchical structure formation scenario based on
the cold dark matter (CDM) model (e.g., Blumenthal et al. 1984), dSphs
are also expected 
to be a key to understand the physical properties of building
blocks. Recently, the observational data for the Local Group dSphs have
become available with relatively high quality and their observed
velocity dispersions imply that dSphs generally have high ratios of
virial mass to luminosity. Although a physical interpretation of this
observational result has long been a matter of debate because tidal
heating by the Galaxy could raise the velocity dispersions (Bellazzini,
Fusi Pecci \& Ferraro 1996; Kroupa 1997; Klessen \& Kroupa 1998), now we
have increasing evidence that dSphs have a large amount of DM (Aaronson
1983; Mateo et al. 1993; Hirashita, Kamaya \& Takeuchi 1999) as
predicted based on the CDM model.

Based on the observational data, Hirashita, Takeuchi, \& Tamura (1998,
hereafter HTT98) have discovered an interesting relation between the
virial mass, $M_{\rm vir}$, and the mass-to-light ratio, $M_{\rm
vir}/L$: there exists a prominent discontinuity at $M_{\rm vir} \sim
10^8 M_\odot$.  Based on the assumption that a dSph is embedded in a
DM-dominated potential well, HTT98 have given a physical interpretation
to this queer relation and suggested a scenario for star formation in a
dSph by considering an initial starburst and a subsequent feedback of
energy from SNe to the ISM. Although their scenario succeeded in
explaining the above relation, other kinds of properties of dSphs must
be compared with it to confirm the interpretation. In particular, their
metallicities should be involved in, since they reflect the star
formation history (e.g., Tinsley 1980). Actually, the chemical
enrichment history of dSphs remains to be tested along with the scenario
in HTT98. Moreover, since the metallicities of the dSphs are very low
([Fe/H] $= -2.5 \sim - 1.5$; Mateo 1998), they provide us valuable
opportunities to study the star formation that occurred in a
low-metallicity gas cloud and to reveal the formation of first luminous
objects in the early universe. In this {\it Letter}, therefore, we focus
on the observed virial mass--metallicity ($M_{\rm vir}$--[Fe/H])
relation, and examine whether it can be consistently understood in the
same theoretical framework that succeeded in explaining the
mass--luminosity ($M_{\rm vir}/L$--$M_{\rm vir}$) relation. The
luminosity--metallicity relation among dSphs as well as dwarf elliptical
galaxies (dEs) near the Galaxy and M31 has been investigated (Caldwell
et al. 1992, 1998; Caldwell 1999; Mateo 1998).  However, if a dSph has a
large amount of DM and formed stars in the DM-dominated potential,
virial mass should be more fundamental than stellar mass in determining
the dynamical timescale.

The rest of this {\it Letter} is organized as follows. In the next
section, we show the observed mass--metallicity relation for the dSphs
after showing the sample and data. In \S~3, the classical understanding
of the relation is presented and the limitations on applying it to the
dSphs are pointed out. In \S~\ref{sec:discuss}, we will propose a new
scenario for chemical enrichment of the dSphs and discuss whether the
mass--metallicity and the mass--luminosity relations are understood
consistently.

\section{THE SAMPLE AND THE MASS--METALLICITY RELATION}

For the two physical parameters, the virial mass ($M_{\rm vir}$) and 
the metallicity (here [Fe/H] is adopted), of the Local Group dSphs, 
we use the values and errors compiled by Mateo (1998).
We adopt the data of Ursa Minor, Draco, Carina, Sextans, Sculptor,
Fornax, Leo I, Leo II, NGC 147, NGC 185, NGC 205 (i.e., the same
sample as that in HTT98). The values are renewed according to
Mateo (1998), but this renewal does not affect the following discussions.
The whole sample is divided into two groups at
$M_{\rm vir} \sim 10^8 M_{\odot}$: the low-mass group includes Ursa
Minor, Draco, Carina, Sextans, Sculptor, Leo I, and Leo II; the others
belong to the high-mass group.
This classification is based on the mass--luminosity relation and is the
same as that adopted by HTT98.

\begin{figure*}
\figurenum{1}
\figcaption[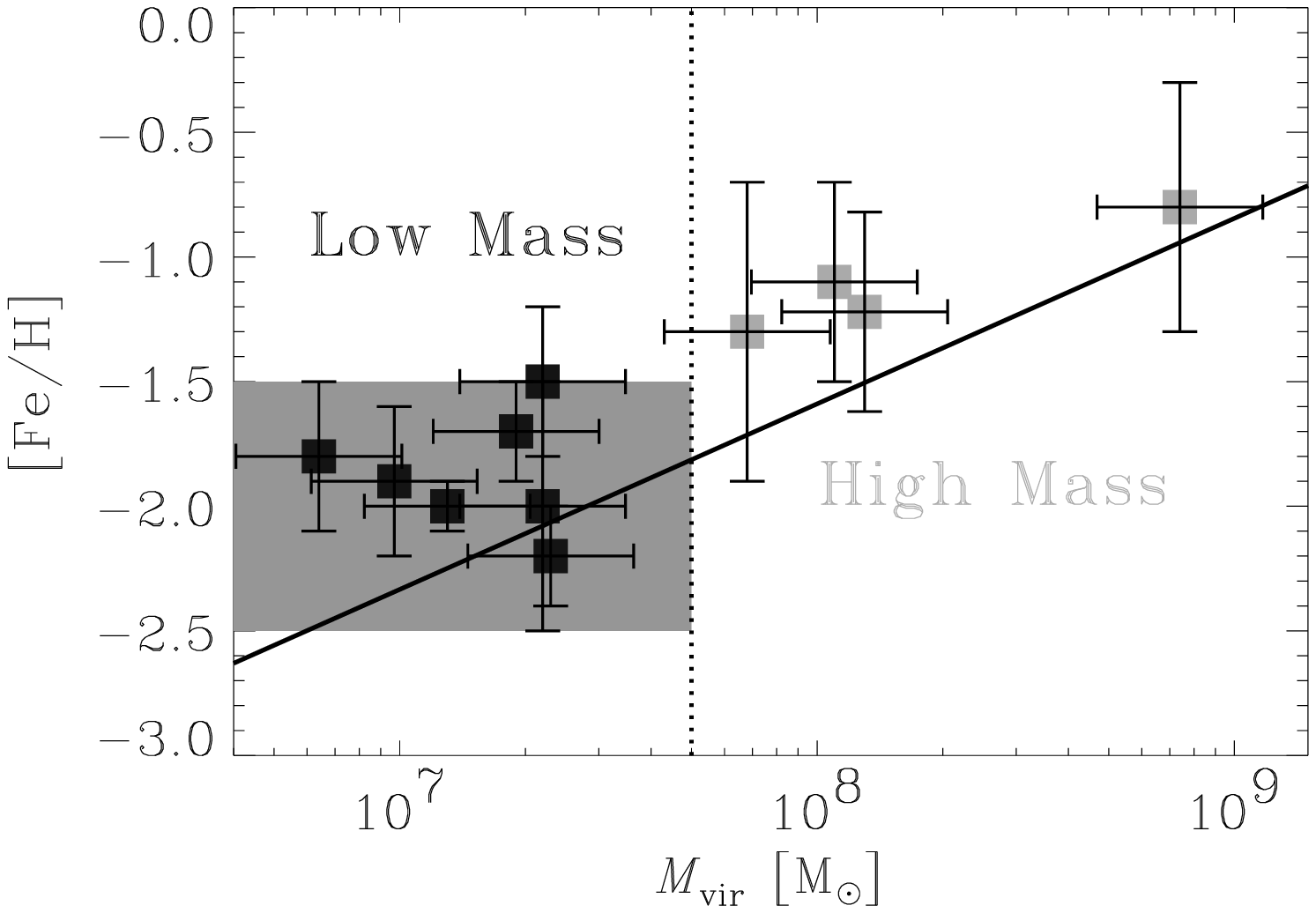]{The relation between the metallicity [Fe/H]
and the virial mass $M_{\rm vir}$ for the Local Group dSphs. 
The dotted vertical line indicates the boundary between the high-mass
dSphs and the low-mass ones (see text). The shaded region shows
an order-of-magnitude spread in metallicity, which is predicted from
the large variety in mass-to-luminosity ratio of the sample.
The solid line represents the model prediction by
YA87.}\label{fig:mass_metal}
\end{figure*}

We show the observational mass--metallicity relation of the dSphs in the
present sample in Figure~\ref{fig:mass_metal}, where a shaded region is
occupied by the low-mass dSphs. As the error bars of [Fe/H] in the
figure, we adopt the intrinsic dispersions of [Fe/H] among stellar
populations of each dSph, which are larger than the mean error of [Fe/H]
(Mateo 1998). For virial masses, errors are calculated by considering
those of the central stellar velocity dispersion of a dSph and the
distance to it from the Galaxy. Since these two errors are on average
$\sim 20$\% and $\sim 10$\%, respectively (Mateo 1998), the resulting
error of a virial mass is evaluated to be a factor of 1.6, which is
adopted in Figure~\ref{fig:mass_metal}. The vertical line ($M_{\rm vir}
= 5 \times 10^7~M_\odot$) indicates the boundary of the two groups. To
include Fornax in the high-mass group according to HTT98, the separating
mass is smaller by a factor of 2 than that in HTT98. In the following,
this separating virial mass ($5 \times 10^7~M_\odot$) is shown as $\sim
10^{8} M_{\odot}$ by adopting its order of magnitude. We believe this
description to be reasonable because it has an error as mentioned above.

The relation between the initial gas mass $M_{\rm G}$ and [Fe/H]
calculated in \citet[hereafter YA87]{yoshii87} is well represented by a
single linear law. The solid line in Figure 1 is obtained by the least
square fitting to the data of Table~1 in YA87, expressed as
\begin{eqnarray}
{\rm [Fe/H]} = 0.745 \log (M_{\rm G}/M_\odot ) - 7.55\, ,
\label{eq:YA_fitting}
\end{eqnarray}
where $M_{\rm G}$ is the initial mass of the system and the
luminosity-weighted [Fe/H] is adopted. In Figure 1, we adopt $M_{\rm G}$
as the virial mass for the convenience in presentation. Since YA87 did
not include the dark matter in their model, we may not be able to
compare this relation directly with our $M_{\rm vir}$--[Fe/H]
relation. However, it can be used as a reference for understanding the
relation.

\section{CLASSICAL PICTURE AND ITS LIMITATION}

\subsection{Binding Energy of the dSphs}

Arimoto \& Yoshii (1987) constructed a chemical and photometric
evolution model for elliptical galaxies (see also Kodama \& Arimoto
1997). Their scenario tells us that a more massive elliptical galaxy can
retain the ISM longer than a less massive one, and will be chemically
enriched further.
Consequently their model predicts the mass--metallicity sequence among
elliptical galaxies, 
and it successfully explains a number of the observational properties of
nearby elliptical galaxies (e.g., Davies et al.\ 1987; Bower, Lucey \&
Ellis 1992) and provides a reasonable picture of their evolutions (e.g.,
Stanford, Eisenhardt, \& Dickinson 1998; Kodama et al. 1998). 
YA87 suggested that the galactic wind model can be extended to the mass
range ($< 10^{9} M_{\odot}$) of dEs and dSphs. The fitting to the
prediction by YA87 (eq.\ [\ref{eq:YA_fitting}]) is shown in
Figure~\ref{fig:mass_metal} ({\it solid line}). It is shown that YA87
roughly reproduced the overall trend in the observed mass--metallicity
relation.

However, the above consideration overlooks a crucial piece in the
physical processes of the galaxy formation, which is the fact that the 
binding energy of a dSph with $M_{\rm vir} \la 10^{8} M_{\odot}$ is
roughly equal to or smaller than the energy released by several SNe (Saito
1979a).
This indicates that once a dSph begins to form stars actively, it can no
longer sustain gas and thus it has no time to proceed with chemical
enrichment efficiently. Therefore, a strongly regulated phase of star
formation with an extremely low rate is necessary to enrich a low-mass
dSph.
Nishi \& Tashiro (2000) have recently shown that a star formation
activity in a cloud whose metallicity is less than 1/100 times the solar
value can be strongly regulated due to the lack of coolant. Since the
observed metallicities of the low-mass dSphs are comparable to 1/100
times the solar value, such a regulation process is likely to have
occurred in their formation epochs.


\subsection{Division of the dSphs into Two Groups at $M_{\rm vir}
\sim10^8 M_{\odot}$}
\label{subsec:htt98}

As considered above, a special care should be taken to understand the
mass--metallicity relation, especially for the dSphs with $M_{\rm vir}
\la 10^{8} M_{\odot}$. Here, we present another piece of evidence that
motivates us to consider that a dSph in the low-mass range must be
separated from a more massive one.

HTT98 have shown that there is a clear difference in the relation
between $M_{\rm vir}/L$ and $M_{\rm vir}$ for the two groups (the
low-mass group with $M_{\rm vir} \la 10^8 M_\odot$ and the high-mass
group with $M_{\rm vir} \ga 10^8 M_\odot$) of the Local Group dSphs. 
The $M_{\rm vir}/L$s of the low-mass dSphs are spread over a
significantly wide range.
HTT98 interpreted this to be a variation of the star formation efficiency
(SFE). Here the SFE indicates the total mass of formed stars divided by
the initial gas mass.
A brief summary of HTT98 will be given in the following.

First, the discontinuity in the $M_{\rm vir}/L$--$M_{\rm vir}$ relation
at $M_{\rm vir} \sim 10^8 M_\odot$ appears at the threshold where the
ISM in the dSphs is blown away efficiently by successive supernovae
(SNe). The virial mass threshold of $10^{8} M_{\odot}$ is derived from
the blow-away condition calculated by Mac Low \& Ferrara (1999). A
proto-dwarf galaxy in the high-mass group would consume gas relatively
completely in making stars because its large binding potential sustains
gas. Its stellar mass is resultantly proportional to initial gas mass
and leads to an almost constant $M_{\rm vir}/L$. Note that the ratio of
initial gas mass and dark mass in a proto-galaxy is assumed to be
constant over entire mass range of dSphs. For a galaxy in the low-mass
group, the ISM is predicted to be blown away efficiently by the SN
explosions after the initial burst of star formation. It is essential to
emphasize here that this scenario is built up on the basis of the fact
that the binding energy of a proto-dwarf galaxy is equal to the energy
released by ten or less SNe.
Therefore, if one OB association forms in a proto-dwarf galaxy, the
galaxy can continue star formation until the heating by SNe affects the
whole galaxy and cause a blowout of the ISM on a short
timescale of $\sim 10^{6}$ yr. 
This implies that the resultant physical parameters of a proto-dwarf
galaxy (mass, metallicity, etc) after a blow-away are determined by the
specific conditions of star formation such as the spatial distribution
of gas, or temperatures of the first OB stars.
As a result, we expect some kinds of stochastic behavior to appear in a
resulting stellar mass and metallicity of the galaxy. Such stochasticity
can actually be found in the $M_{\rm vir}/L$--$M_{\rm vir}$ relation of
the low-mass dSphs.
According to Figure~1 in HTT98, the mass-to-light ratio ($M_{\rm
vir}/L$) ranges from $\sim 10$ to $\sim 100$ with an order-of-magnitude
scatter.\footnote{Although HTT98 used a mass-to-light ratio of Leo~I of
$\sim 1$, \citet{mateo98} showed that it is 4.6. We use the latter value
here.}
As explained in \S~\ref{subsec:origin}, this is suggested to be caused
by a scatter of SFE with a similar order of magnitude.

Finally, it is worth mentioning that the high-mass dSphs are also
located on the extension from the sequence of dEs in the
mass--luminosity relation as well as the mass--metallicity relation.
Actually, Mac Low \& Ferrara (1999) suggested that the ISM in a
high-mass dSph is not easily blown away by an initial burst, which
implies that the classical picture holds for the high-mass dSphs.
In contrast, it is suggested to be difficult for the low-mass dSphs to
maintain star formation against the stellar feedback to proceed with 
chemical enrichment to the observed level because of their too small 
binding energy.
Therefore, in the following, we focus on the dSphs in the low-mass group
as a target to reconsider how their chemical enrichment has proceeded.

\section{DISCUSSIONS}\label{sec:discuss}

\subsection{What Determines the Metallicity Level of the Low-Mass dSphs ?}

If a formation of a dSph started from a proto-galactic gas cloud, the
subsequent star formation after a massive star forms is suggested to be
self-regulated as long as the metallicity is $\la 1/100$ times the solar
value (Nishi \& Tashiro 2000).
This is explained as follows. Since the metal-line cooling is extremely
inefficient in such a low-metallicity environment, the coolant is ${\rm
H}_2$.
However, the UV radiation from a central massive star dissociates the
H$_2$ molecules, causing a strong regulation of star formation.
This process avoids forming stars efficiently in a dSph belonging to the
low-mass group and enables a dSph to continue star formation without
blowing away the ISM and allow chemical enrichment to proceed. Since
several SNe release energy comparable to the binding energy of the ISM,
a star formation rate of a dSph in this phase should be considerably low
and thus chemical enrichment would also proceed gradually. If the
metallicity finally reaches to $\mbox{[Fe/H]} \sim -2$, with which the
metal cooling becomes effective, a starburst may occur and a following
galactic wind would blow away the ISM. Nishi \& Tashiro (2000) discussed
how many SNe are required for a primordial gas cloud with $\sim 10^6
M_{\odot}$ to have $\mbox{[Fe/H]} \sim -2$. If their discussion is
extended to the mass range for the dSphs in the low-mass group, it is
found that about 500 SNe are necessary, and thus a timescale of $\sim
1$~Gyr is required. Here we assumed that SN explosions caused by a
massive star whose lifetime is $\sim 10^6$ yr subsequently occurs
one-by-one in a proto-galactic gas cloud with the gas mass of $10^7
M_{\odot}$.

The mass of stars formed in the regulation phase can be estimated as
follows. Adopting the Salpeter's initial mass function (IMF) with the
lower and upper cutoffs $0.1M_\odot$ and $100M_\odot$, respectively, it
is shown that the averaged mass of SN progenitors (stars whose mass is
larger than $10M_\odot$) is $20M_\odot$.
The total mass of the high-mass stars (i.e., SN progenitors) is
$10^4M_\odot$. Applying the same IMF, the total stellar mass is
calculated to be $8\times 10^4M_\odot$. This is clearly much smaller
than the present stellar mass of dSphs. Thus, the majority of stars in
the dSphs must be formed in the starburst after the regulation becomes
ineffective.
Consequently, if further significant chemical enrichment does not occur
because of an instantaneous blow-out after the burst, a dSph with
$M_{\rm vir} \la 10^8 M_{\odot}$ would have an [Fe/H] $\sim -2$ at the
present epoch, which is consistent with the observed metallicity as
shown in Figure~1.

\subsection{Origin of the Scatter in the Mass--Metallicity Relation
among the Low-Mass dSphs}\label{subsec:origin}

In the previous subsection, the origin of the observed metallicity
levels of the dSphs in the low-mass group was explored. As shown in
Figure 1, it should also be emphasized that the mass--metallicity
relation among the low-mass dSphs is characterized by a significant
scatter of metallicity. We are now in a position to consider the origin
of this metallicity variation.

According to Pagel (1997, p.\ 219), the metallicity ($Z$) of a galaxy as
a closed system is expressed as
\begin{eqnarray}
Z=\left. p_{\rm eff}\,\ln\frac{s+g}{g}\right|_{t=t_{\rm gw}},
\end{eqnarray}
where $p_{\rm eff}$ is the yield of the chemical enrichment effective in
the lifetime of the galaxy, and $s$ and $g$ are the masses of the stars
and the gas, respectively.
The metallicity is evaluated at the galactic wind era, $t_{\rm gw}$,
since the chemical enrichment of the interstellar gas can proceed only
up to $t_{\rm gw}$.
Since dSphs lost a significant fraction of their gas at $t_{\rm gw}$
(Saito 1979b), $s\ll g$ is adopted.
Then we obtain
\begin{eqnarray}
  Z\simeq\left. p_{\rm eff}\,\frac{s}{g}\right|_{t=t_{\rm gw}}
  \equiv p_{\rm eff}\,{\rm SFE}\, ,\label{eq:sfe}
\end{eqnarray}
where we have defined the star formation efficiency (SFE) as $s/g$ at
$t_{\rm gw}$, that is, the stellar mass normalized by the gas mass at
$t_{\rm gw}$.

As stated in \S~3.2, if the initial gas mass is proportional to the
virial mass, $M_{\rm vir}/L$ of a dSph in the low-mass group is
approximately proportional to 1/SFE. Following this context, the SFEs of
the dSphs in the low-mass group are expected to spread over an order of
magnitude because their $M_{\rm vir}/L$s are distributed over a wide
range with the similar magnitude.
Since we see from equation (\ref{eq:sfe}) that the metallicity varies
with SFE and, indeed, the [Fe/H] for the low-mass dSphs has about 1
order of magnitude scatter, it is suggested that the scatter of [Fe/H]
in the low-mass dSphs is consistent with the variation of $M_{\rm
vir}/L$.

Finally, we discuss the physical origin of the scatter of SFE among the
low-mass dSphs. As mentioned in \S~\ref{subsec:htt98}, this is likely to
originate from that of initial condition in the formation of low-mass
dSphs. Mac Low \& Ferrara (1999) suggested that a blow-away condition
depends on the ellipticity of initial gas density distribution. This
indicates that a star formation in a proto-dwarf galaxy also depends on
the initial ellipticity, which is presumably determined at random (e.g.\
Bardeen et al. 1986). 
One of the other possibilities for the scatter of SFE is a variation in
temperatures of OB stars formed in a starburst.  Since their ionizing
photons dissociate a surrounding H~{\sc i} region and suppress star
formation, an SFE in a forming dSph is considered to be affected by
their temperatures. For a low-mass dSph, this effect can be severe
because only a small number of OB stars form at one time and thus
temperatures that he stars have are considered to be determined
stochastically. 
In order to evaluate quantitatively to what extent the observed scatter
is reproduced by a reasonable range of variations of initial conditions,
further theoretical studies with stochastic processes are needed. It
might be found from such studies that some external factors like
environmental effects are required.

It may be noticed that in some dSphs (e.g., Draco, Andromeda II), large
internal dispersions of metallicity among the stars are observed. 
If we consider the inhomogeneity of the distribution of gas density in a
proto-galactic cloud, chemical enrichment in the galaxy can also proceed
inhomogeneously because the outflow efficiency of gas differs from place
to place. As a result, internal dispersions of metallicity can be
produced.
Such an inhomogeneity could affect a measurement of an average [Fe/H] in
a dSph because [Fe/H] of each dSph has been determined mainly by
examining only a small number of giant stars especially for the Local
Group dSphs. It may be hence implied that an intrinsic dispersion of
metallicity among the dSphs is difficult to find based on the observed
scatter of [Fe/H].

\acknowledgements 
We are grateful to the anonymous referee for useful comments to improve
the paper. We are grateful to R. Nishi and M. Tashiro for discussions
and K. Ohta, S. Mineshige, and H. Shibai for continuous encouragement.
One of us (H.H.) acknowledges the Research Fellowship of the Japan
Society for the Promotion of Science for Young Scientists.



\end{document}